%% file: weaver8.tex
\documentclass[aps,pra,twocolumn]{revtex4}
\usepackage{amsmath}
\usepackage{amssymb}

\usepackage{epsf,graphicx,rotating,color}

\newcommand{\rmd}{{\rm d}}

\newcommand{\eps}{\varepsilon}

\newcommand{\la}{\langle}
\newcommand{\ra}{\rangle}

\begin{document}

\title{Scattering fidelity in elastodynamics}

\author{T. Gorin}
\affiliation{Max-Planck-Institut f\" ur Physik komplexer Systeme, 
        N\" othnitzer Str. 38, D-01187 Dresden, Germany}

\author{T. H. Seligman}
\affiliation{Centro de Ciencias F\'{\i}sicas, Universidad Nacional
Aut\'{o}noma de M\'{e}xico, Campus Morelos, C.~P. 62251,
Cuernavaca, Morelos, M\'{e}xico}

\author{R. L. Weaver}
\affiliation{Theoretical and Applied Mechanics, University of Illinois, 
104 South Wright Street, Urbana, Illinois 61801, USA}
 

\begin{abstract}
The recent introduction of the concept of scattering fidelity, causes us to 
revisit the experiment by Lobkis and Weaver 
[Phys. Rev. Lett. {\bf 90}, 254302 (2003)].
There, the ``distortion'' of the coda of an acoustic signal is measured
under temperature changes. This quantity is in fact the negative logarithm 
of scattering fidelity. We re-analyse their experimental data for two samples, 
and we find
good agreement with random matrix predictions for the standard fidelity. 
Usually, one may expect such an agreement for chaotic systems only.
While the first sample, may indeed be assumed chaotic, for the second sample,
a perfect cuboid, such an agreement is more surprising. For the first sample, 
the random matrix analysis yields a perturbation strength compatible with 
semiclassical predictions. For the cuboid the measured perturbation strength 
is much larger than expected, but with the fitted values for this strength,
the experimental data are well reproduced.
\end{abstract}

\maketitle

Lobkis and Weaver (henceforth LW) experiment~\cite{LobWea03} 
have measured  the sensitivity of elastic coda waves to temperature changes. 
After correcting for a trivial term due to change of volume and wavespeed 
they quantify the 
effect as ``distortion'', and study its decay as a function of time. We shall 
show that this quantity is the negative logarithm of the ``scattering 
fidelity'' introduced and measured in~\cite{SSGS04,SGSS05}. For sufficiently 
chaotic dynamics in systems weakly coupled to decay channels, scattering 
fidelity approaches the standard fidelity amplitude. Fidelity, which is 
the absolute value squared of the fidelity amplitude, has received a great 
deal of attention in recent years, as it is used as a benchmark in quantum 
information processes~\cite{NieChu00}, and in the context of 
quantum chaos (for a partial overview and references see~\cite{PSZ03}). 

For the above mentioned reasons, we expect the random matrix analysis of 
fidelity decay~\cite{GPS04}, to apply to the experiments of LW, as it does in 
the case of comparable experiments with chaotic microwave 
cavities~\cite{SSGS04,SGSS05}. Random matrix theory (RMT) makes a definite 
statement on the form of the fidelity amplitude as a function of time. It 
yields a unified description of the ``perturbative'' and the ``Fermi golden
rule''~\cite{Jac01,Cuc02} regime. This analysis is more appropriate than the 
treatment in~\cite{LobWea03}, which seems to be on a footing similar to the
Fermi golden rule results obtained in~\cite{Cuc02}. RMT still contains the 
perturbation strength as a free parameter, which must be determined 
independently. For quantum chaotic systems, different semiclassical methods 
have been used~\cite{LebSie99,Cuc02,SGSS05}. It is still an open problem to
extend those results to the perturbative regime~\cite{SGSS05}.

We analyzed the data for two samples, measured by LW, both being
three dimensional objects of similar size in all three spatial directions. The
first object is the so called ``medium block'' (after extra cut) which is 
supposed to have dominantly chaotic dynamics, with no symmetries left. The 
second object is a cuboid, called ``rectangle'', where the dynamics is
not chaotic, but due to mode conversion it is also not integrable and actually 
known to display random matrix behavior~\cite{Schaa03}. In both cases,
we obtained very good agreement with our RMT result, as far as the form
of the scattering fidelity functions is concerned. For the first sample, we 
also obtain a reasonable agreement for the perturbation strength (comparing
the fitted values from an RMT analysis with the theoretical result of LW).
For the second sample, we obtain values which are systematically too large,
by a factor of two, approximately.
Over all, we find that our random matrix approach provides a very good 
framework, for the discussion of distortion measurements of non-integrable 
elastodynamic systems, thus providing additional information for the elastic 
problem. Conversely, experimental results with adequate control of the 
perturbation strength are rare in fidelity analysis and thus the possibility 
to use acoustic scattering results is highly welcome.

\paragraph*{Scattering fidelity}
The scattering fidelity \cite{SGSS05} is defined as
\begin{equation}
f_{ab}(t)= \la \hat S_{ab}(t)^*\, \hat S'_{ab}(t)\ra / \sqrt{
   \la |\hat S_{ab}(t)|^2\ra \, \la |\hat S'_{ab}(t)|^2\ra } \; .
\label{scattfid}\end{equation}
Here, $\la \hat S_{ab}(t)^*\, \hat S'_{ab}(t)\ra$ is the Fourier transform 
of the cross-correlation function of a scattering matrix element
for the unperturbed and perturbed system, respectively. Similarly,
$\la |\hat S_{ab}(t)|^2\ra$ and $\la |\hat S'_{ab}(t)|^2\ra$ are the Fourier 
transforms of the corresponding autocorrelation functions, used for proper 
normalization. It can be shown that appropriate averaging (denoted by 
$\la\ldots\ra$) yields the standard fidelity amplitude for chaotic systems 
weakly coupled to the scattering channels~\cite{SSGS04}. In that case, a RMT
model can be used to compute the fidelity amplitude $f(t)$ 
on the basis of linear response theory~\cite{GPS04}. With $x= t/t_H$ ($t_H$ is 
the Heisenberg time) and $b_2(x)$, the two point form factor for the Gaussian
orthogonal ensemble, one obtains:
\begin{equation}
-\ln f(t) = \lambda_0^2\! \left[ \frac{x}{2}\! +\! x^2\! - 
   \!\int_0^x\!\!\rmd x'\!\!\int_0^{x'}\!\!\rmd x'' \; b_2(x'') 
   \right] .
\label{flresp}\end{equation}
Here, the ``perturbative'' and the ``Fermi golden rule''
regime arise as two particular limits:
For small $\lambda_0$ the decay of $f(t)$ is dominantly Gaussian, while for
large $\lambda_0$ it is dominantly exponential~\cite{GPS04}. In a microwave
experiment excellent agreement between $f_{ab}(t)$ and $f(t)$ has been 
found~\cite{SSGS04,SGSS05}. The validity of Eq.~(\ref{flresp}) for the
fidelity of closed chaotic systems has been demonstrated with numerical 
calculations for the kicked rotor~\cite{HB05}.

In~\cite{LobWea03} the authors measure the acoustic response to a short 
piezoelectric pulse as a function of time. They consider the normalized
cross correlation between two such signals obtained at different temperatures
$T_1$ and $T_2$:
\begin{equation}
X(\varepsilon)= \frac{\int\rmd t\; S_{T_1}(t)\; S_{T_2}(t(1+\eps))}
   {\sqrt{\int\rmd t\; S_{T_1}^2(t)\; \int\rmd t\; S_{T_2}^2(t(1+\eps))}} \; .
\label{Xeq}\end{equation}
This equation displays a very similar structure as Eq.~(\ref{scattfid}). The 
time averaging over a small window, performed in Eq.~(\ref{Xeq}), corresponds 
to a smoothing of the correlation functions in Eq.~(\ref{scattfid}). 
The selection of  $\varepsilon$, such that the correlation function 
$X(\varepsilon) = X_{\rm max}$ becomes maximal, eliminates the 
trivial effects due to dilation and change of wavespeed, caused by the temperature
change. This is 
quite similar to the 
spectral unfolding performed in Ref.~\cite{SSGS04} to eliminate the 
volume effect of a moving wall. The distortion is defined as 
$D(t) = -\ln (X_{\rm max})$, where the time dependence is given by the 
``age'' of the signal, {\it i.e.} the center of the small time-interval over 
which the correlation function $X(\varepsilon)$ was evaluated. If formulated 
as a scattering process, we find $D(t) = -\ln[f_{aa}(t)]$, 
where the scattering channel $a$ is defined by the transducer, that transmits 
excitation to and from the sample. Thus, for sufficiently chaotic samples in 
the elastodynamic scattering experiments,
we expect that the scattering fidelity is equal to the fidelity amplitude,
and that the latter is well described by the RMT result, Eq.~(\ref{flresp}).

In~\cite{LobWea03} the perturbation strength is estimated in terms of the
distortion coefficient 
$C \approx 3.26\times 10^{-10}/({\rm K}^2\, {\rm cm}\, {\rm msec}^{-1})$. This 
value is based on a ray picture of the resonating acoustic waves, the
assumption of random reflection angles along the ray paths, and an estimate 
for the mode conversion rates between dilational and shear waves.
We use that value to obtain an estimate for the dimensionless 
perturbation strength $\lambda_0$ in Eq.(\ref{flresp}):
\begin{equation}
\lambda_0 = f\; \sqrt{2\, c\; t_H(f)}\qquad
c= C\; \Delta^2\; V/S \; ,
\label{laestim}\end{equation}
where $\Delta$ denotes the temperature difference $T_1-T_2$ in Kelvin, while
$V$ and $S$ denote the volume and the surface of the sample. From 
Refs.~\cite{Dup60,Saf92}, we obtain the following expression for the 
Heisenberg time $t_H$:
\begin{equation}
t_H = \frac{4\pi}{c_p^3}\; V\; (2q^3+1)\; f^2 +
   \frac{\pi}{2\, c_p^2}\; S\left( 3q + \frac{2}{q^2 - 1} \right) f \; .
\label{heist}\end{equation}
Here, the longitudinal wave velocity is $c_p= 637\,$cm/msec and the transverse 
(shear) wave velocity is $c_s= c_P/q = 316\,$cm/msec. The frequency (range) is 
denoted by $f$.

\begin{figure}[t]
\input{dist}
\caption{(Color online)
The distortion for the medium block as plotted in Fig.~4 of 
Ref.~\cite{LobWea03}. The thin jagged lines correspond to measurements in the 
frequency ranges $300\,$kHz, $600\,$kHz, $700\,$kHz, and $800\,$kHz (from 
bottom to top). The thick lines show $-\ln f(t)$ according to
Eq.~(\ref{flresp}), using the values of $\lambda_0$, as given in Tab.~1
(full RMT fit). For the thin straight lines, only the linear term in 
Eq.~(\ref{flresp}) has been taken into account.}
\end{figure}
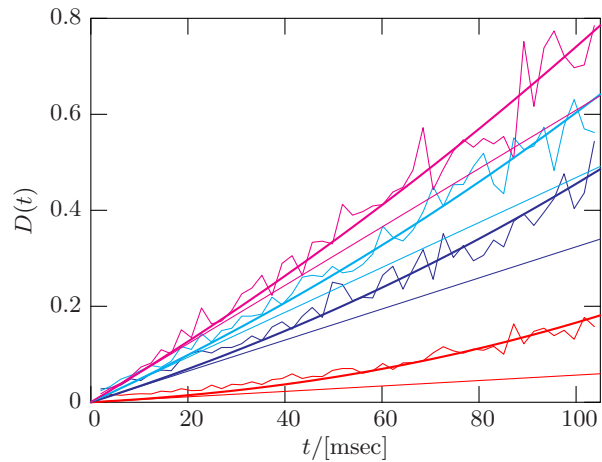

\paragraph*{Medium block}
That specimen has volume $V= 906\, $cm$^3$ and 
$S= 636\, $cm$^2$, and the temperature difference was $\Delta= 4$K. The 
distortion, measured as a function of time is shown in 
Fig.~1. With the Heisenberg times given in Eq.~(\ref{heist}), we may fit the
perturbation strength $\lambda_0$ and compare to the estimate
in Eq.~(\ref{laestim}). For experimental reasons, the data for $D(t)$ is 
reliable for $t> 20\, $msec, only. The fits, mentioned above, have been 
restricted correspondingly.

The agreement between the measured distortions or scattering fidelities and 
random matrix theory is within the statistical error of the data.
A deviation from linear behavior is hardly noticeable, except for the 
$300\, $kHz data, which is the only case where 
$t_H\approx 78\, $msec~\cite{LWcomm} lies 
within the time range of the figure. However, there is still a considerable 
effect on the fit values for $\lambda_0$. This can be seen from the dashed
curves, which have been obtained by taking only the linear part of 
Eq.~(2) into account.

\begin{table}
\begin{tabular}{l|clll}
 $f/$kHz & $\lambda_0$ & (lin. fit) & (full fit) & $t_H/{\rm msec}$\\
\hline
300 & & 0.471 & 0.298 &  78.4\\
600 & & 1.584 & 1.381 & 294.6\\
700 & & 2.136 & 1.929 & 397.3\\
800 & & 2.709 & 2.505 & 515.3
\end{tabular}
\caption{Medium block: Table of linear and full RMT fit values for the 
dimensionless perturbation strength $\lambda_0$ together with the respective 
Heisenberg times for the four frequency ranges, analysed.}
\end{table}

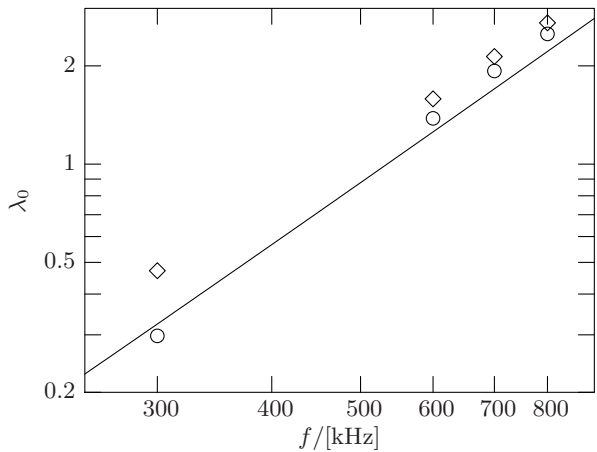
\begin{figure}[t]
\input{pstrength}
\caption{Medium block:
The perturbation strength as a function of the frequency range. The circles
show the values for $\lambda_0$ obtained from a fit with Eq.~(\ref{flresp}), 
while the diamonds show the corresponding values when only the linear term
of Eq.~(\ref{flresp}) is taken into account. The solid line gives the 
perturbation strength as obtained from Eq.~(\ref{laestim}).}
\end{figure}

Tab.~1 gives the values of $\lambda_0$ as obtained from linear fits to the 
data (such a procedure has been applied in~\cite{LobWea03}), and from fits 
with the full random matrix result, Eq.~(\ref{flresp}). In the last column, 
the respective Heisenberg times are given.
In Fig.~2, the fitted perturbation strengths are plotted versus frequency. 
The solid line shows the estimated behavior of the perturbation strength, 
computed from Eq.~(\ref{laestim}). As long as the volume
term dominates in the expression~(\ref{heist}) for $t_H$, the perturbation 
strength increases quadratically with the frequency. 
The difference between the linear fits (diamonds) and the full RMT fits 
(circles) is clearly noticeable. The full
RMT analysis moves the fit values for $\lambda_0$ quite close to the 
prediction of LW. We have no explanation 
for the remaining deviations. It is not clear, whether these are
statistically acceptable, whether chaoticity is not perfect, or wether 
there is some other reason. We may recall that in Ref.~\cite{SGSS05},
the experimental perturbation strength did also not agree with the theoretical 
estimate, as it was too small. However, there, the perturbation 
strength could be measured independently, from the level dynamics. This
demonstrated that for the cavities under study, the semiclassical 
approximation was not yet justified. It would be desirable, if the level 
dynamics could be measured in the context of the present experiment.

\begin{table}
\begin{tabular}{l|clll}
 $f/$kHz & $\lambda_0$ & (lin. fit) & (full fit) & $t_H/{\rm msec}$\\
\hline
100 & & 0.241 & 0.061 & 10.36\\
200 & & 0.607 & 0.279 & 35.75\\
300 & & 0.903 & 0.567 & 76.19\\
400 & & 1.389 & 1.033 & 131.68\\
500 & & 1.905 & 1.561 & 202.21\\
600 & & 2.564 & 2.227 & 287.78\\
700 & & 3.282 & 2.957 & 388.39\\
800 & & 3.969 & 3.662 & 504.04
\end{tabular}
\caption{Rectangle: Table of linear and full RMT fit values for the 
dimensionless perturbation strength $\lambda_0$ together with the respective 
Heisenberg times.}
\end{table}

\begin{figure}[t]
\input{distrect}
\caption{(Color online) 
The distortion as a function of time, for the rectangle. The thin jagged lines 
correspond to measurements in the frequency ranges from $100\, $kHz to  
$800\, $kHz, in steps of $100\, $kHz (from bottom to top). The thick smooth 
lines show the best fits with $-\ln f(t)$, according to Eq.~(2), with 
$\lambda_0$ given in Tab.~2 (full RMT fit).}
\end{figure}
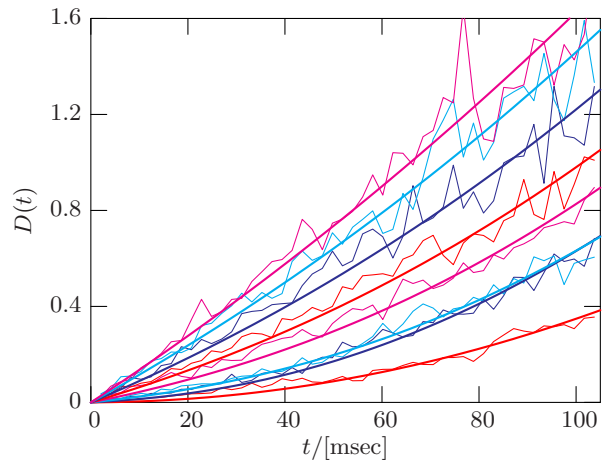

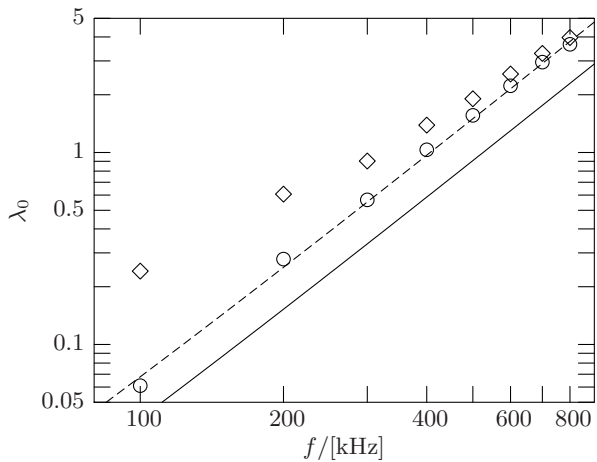
\begin{figure}[t]
\input{plamrect}
\caption{Rectangle:
The perturbation strength as a function of the frequency range. The circles
show the values for $\lambda_0$ as obtained from Eq.~(\ref{flresp}),
and from the same expression, taking into account only the linear term 
(diamonds). The solid line gives the perturbation strength as given by
Eq.~(\ref{laestim}). The dashed line shows the same expression, but with 
$C= 8.90\times 10^{-10}/({\rm K}^2\, {\rm cm}\, {\rm msec}^{-1}) $ obtained 
from a fit to the data points (circles).}
\end{figure}

\paragraph*{Rectangle}
In Fig.~3 and Fig.~4, we analyse data for the rectangle, a perfect cuboid.
Clearly,
a scalar wave equation would lead to integrable ray dynamics, where our random 
matrix model must fail. However, in the present case, the wave field has two
components (dilatational and a shear waves), which are coupled due to mode 
conversion. The corresponding classical dynamics are marginally stable, but
may still be ergodic. Recently, the elastodynamic system has been studied 
thoroughly, with a focus on spectral statistics~\cite{Schaa03}. Apart from
certain symmetries, the statistical measures show accurate RMT behavior.
Nevertheless, from a theoretical point of view, it remains 
an open question, whether scattering 
fidelity still agrees with standard fidelity, and whether the behavior of 
the fidelity amplitude can be described by random matrix theory. The 
following analysis with Eq.~(\ref{flresp}) will shed some light on these 
questions.

For the rectangle, we have data for eight
frequency windows at $f= 100\,$kHz, $200\,$kHz, \ldots, $800\,$kHz.
The corresponding Heisenberg times range from $10\,$msec to $500\,$msec 
(see Tab.~2). In Fig.~3, one can clearly see a transition from linear to 
quadratic decay, characteristic for the RMT expression, Eq.~(\ref{flresp}). 
Here, it is really surprising that the RMT expression describes the data so 
well. The perturbation strengths obtained from fits to the RMT expression
are plotted in Fig.~4 (circles). Except for a constant factor (the distortion 
coefficient), the result follows the theoretical expectation
(Eq.~(\ref{laestim}), solid line). A fit 
for the distortion coefficient on the basis of Eq.~(\ref{laestim}) yields the 
dashed line. The values for $\lambda_0$ obtained from a linear fit (diamonds)
clearly, do not follow the theory.

In this work, we identified a previously published elastomechanic scattering 
experiment as an experiment that measures scattering fidelity in a setting 
usualy called echo dynamics. As in the case of electromagnetic billiards, 
it turns out that 
a simple random matrix model describes the decay of the scattering fidelity
very well, despite of the fact that these are much more complicted systems. 
A number of questions could not be answered. 
In the case of the medium block, we had not enough data, in the low frequency
region, where the crossover between Fermi golden rule and perturbative regime
occurs. Also, we could not consider any possible symmetries of the system,
which might be one source for the remaining differences between 
estimated and measured perturbation strengths. 
For the rectangle, our results complement those of ~\cite{Schaa03}. They show
that at least in one important aspect the wave functions behave like those 
of a chaotic system.

The high quality factors of elastomechanic experiments, as well as the 
possibility to measure explicitely in the time domain, make these experiments 
particularly atractive. Among possible experiments, we believe that it will be 
worthwhile to analyze strong perturbation data, {\it e.g.} larger temperature 
differences. Such experiments are more difficult, due to precision problems, but 
may serve to explore the limits of RMT, or detect other regimes of fidelity decay.
This is particularly interesting, because 
the exact solution of the RMT model is now available~\cite{SS04}.

We wish to thank O. I. Lobkis for discussions and for data and results put 
at our disposition. T.H.S. acknowledges financial support under 
grants CONACyT \# 41000-F and UNAM-DGAPA IN-101603. R.L.W. acknowledges support
from NSF CMS-0201346.

%

\end{document}

%% file: dist.tex
\begin{picture}(0,0)%
\includegraphics{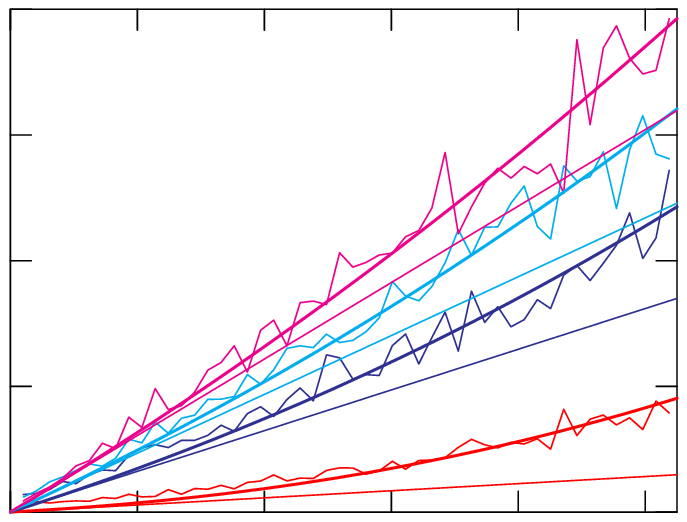}%
\end{picture}%
\setlength{\unitlength}{0.0200bp}%
\begin{picture}(11699,8640)(0,0)%
\put(1400,1050){\makebox(0,0)[r]{\strut{} 0}}%
\put(1400,2860){\makebox(0,0)[r]{\strut{} 0.2}}%
\put(1400,4670){\makebox(0,0)[r]{\strut{} 0.4}}%
\put(1400,6480){\makebox(0,0)[r]{\strut{} 0.6}}%
\put(1400,8290){\makebox(0,0)[r]{\strut{} 0.8}}%
\put(1575,700){\makebox(0,0){\strut{} 0}}%
\put(3404,700){\makebox(0,0){\strut{} 20}}%
\put(5232,700){\makebox(0,0){\strut{} 40}}%
\put(7061,700){\makebox(0,0){\strut{} 60}}%
\put(8889,700){\makebox(0,0){\strut{} 80}}%
\put(10718,700){\makebox(0,0){\strut{} 100}}%
\put(350,4670){\rotatebox{90}{\makebox(0,0){\strut{}$D(t)$}}}%
\put(6375,175){\makebox(0,0){\strut{}$t/[{\rm msec}]$}}%
\end{picture}%
 

%% file: pstrength.tex
\begin{picture}(0,0)%
\includegraphics{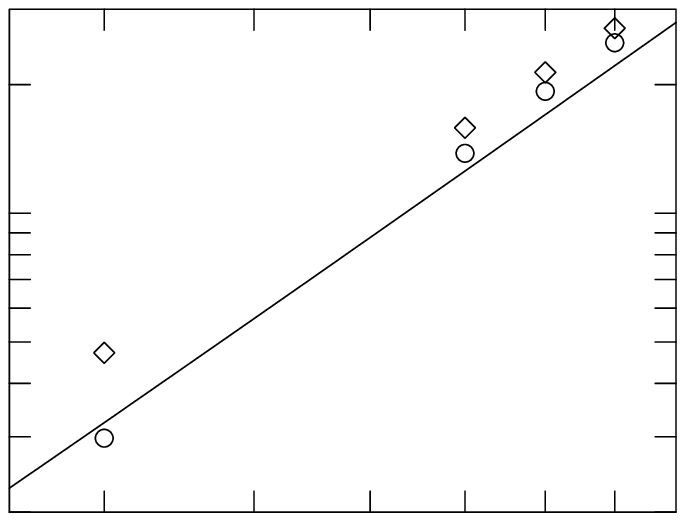}%
\end{picture}%
\setlength{\unitlength}{0.0200bp}%
\begin{picture}(11699,8640)(0,0)%
\put(1400,1050){\makebox(0,0)[r]{\strut{} 0.2}}%
\put(1400,2134){\makebox(0,0)[r]{\strut{}}}%
\put(1400,2903){\makebox(0,0)[r]{\strut{}}}%
\put(1400,3500){\makebox(0,0)[r]{\strut{} 0.5}}%
\put(1400,3987){\makebox(0,0)[r]{\strut{}}}%
\put(1400,4399){\makebox(0,0)[r]{\strut{}}}%
\put(1400,4756){\makebox(0,0)[r]{\strut{}}}%
\put(1400,5071){\makebox(0,0)[r]{\strut{}}}%
\put(1400,5353){\makebox(0,0)[r]{\strut{} 1}}%
\put(1400,7206){\makebox(0,0)[r]{\strut{} 2}}%
\put(2941,700){\makebox(0,0){\strut{} 300}}%
\put(5097,700){\makebox(0,0){\strut{} 400}}%
\put(6770,700){\makebox(0,0){\strut{} 500}}%
\put(8136,700){\makebox(0,0){\strut{} 600}}%
\put(9292,700){\makebox(0,0){\strut{} 700}}%
\put(10292,700){\makebox(0,0){\strut{} 800}}%
\put(350,4670){\rotatebox{90}{\makebox(0,0){\strut{}$\lambda_0$}}}%
\put(6375,175){\makebox(0,0){\strut{}$f/[{\rm kHz}]$}}%
\end{picture}%
 

%% file: distrect.tex
\begin{picture}(0,0)%
\includegraphics{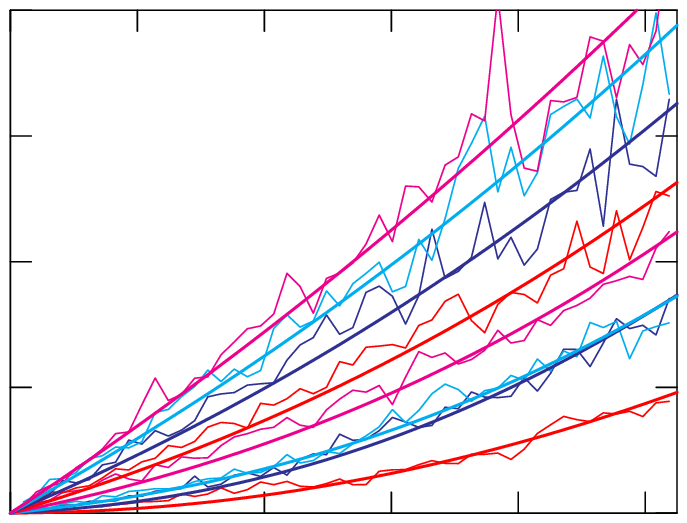}%
\end{picture}%
\setlength{\unitlength}{0.0200bp}%
\begin{picture}(11699,8640)(0,0)%
\put(1400,1050){\makebox(0,0)[r]{\strut{} 0}}%
\put(1400,2860){\makebox(0,0)[r]{\strut{} 0.4}}%
\put(1400,4670){\makebox(0,0)[r]{\strut{} 0.8}}%
\put(1400,6480){\makebox(0,0)[r]{\strut{} 1.2}}%
\put(1400,8290){\makebox(0,0)[r]{\strut{} 1.6}}%
\put(1575,700){\makebox(0,0){\strut{} 0}}%
\put(3404,700){\makebox(0,0){\strut{} 20}}%
\put(5232,700){\makebox(0,0){\strut{} 40}}%
\put(7061,700){\makebox(0,0){\strut{} 60}}%
\put(8889,700){\makebox(0,0){\strut{} 80}}%
\put(10718,700){\makebox(0,0){\strut{} 100}}%
\put(350,4670){\rotatebox{90}{\makebox(0,0){\strut{}$D(t)$}}}%
\put(6375,175){\makebox(0,0){\strut{}$t/[{\rm msec}]$}}%
\end{picture}%
 

%% file: plamrect.tex
\begin{picture}(0,0)%
\includegraphics{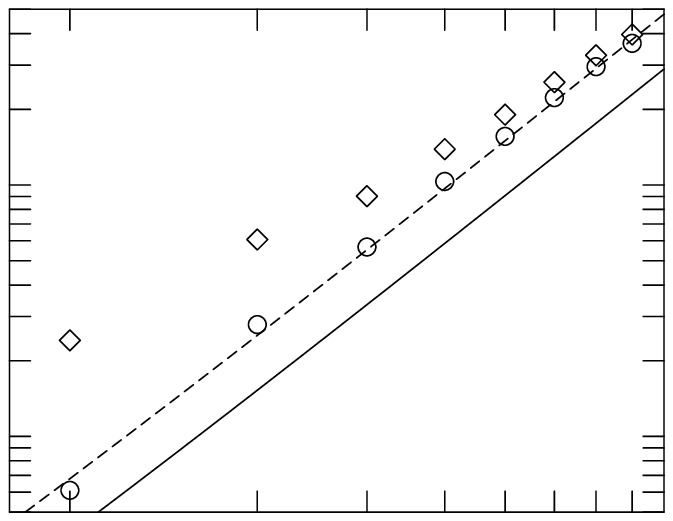}%
\end{picture}%
\setlength{\unitlength}{0.0200bp}%
\begin{picture}(11699,8640)(0,0)%
\put(1575,1050){\makebox(0,0)[r]{\strut{} 0.05}}%
\put(1575,1337){\makebox(0,0)[r]{\strut{}}}%
\put(1575,1579){\makebox(0,0)[r]{\strut{}}}%
\put(1575,1789){\makebox(0,0)[r]{\strut{}}}%
\put(1575,1974){\makebox(0,0)[r]{\strut{}}}%
\put(1575,2140){\makebox(0,0)[r]{\strut{} 0.1}}%
\put(1575,3229){\makebox(0,0)[r]{\strut{}}}%
\put(1575,3867){\makebox(0,0)[r]{\strut{}}}%
\put(1575,4319){\makebox(0,0)[r]{\strut{}}}%
\put(1575,4670){\makebox(0,0)[r]{\strut{} 0.5}}%
\put(1575,4957){\makebox(0,0)[r]{\strut{}}}%
\put(1575,5199){\makebox(0,0)[r]{\strut{}}}%
\put(1575,5409){\makebox(0,0)[r]{\strut{}}}%
\put(1575,5594){\makebox(0,0)[r]{\strut{}}}%
\put(1575,5760){\makebox(0,0)[r]{\strut{} 1}}%
\put(1575,6849){\makebox(0,0)[r]{\strut{}}}%
\put(1575,7487){\makebox(0,0)[r]{\strut{}}}%
\put(1575,7939){\makebox(0,0)[r]{\strut{}}}%
\put(1575,8290){\makebox(0,0)[r]{\strut{} 5}}%
\put(2619,700){\makebox(0,0){\strut{} 100}}%
\put(5318,700){\makebox(0,0){\strut{} 200}}%
\put(6897,700){\makebox(0,0){\strut{}}}%
\put(8017,700){\makebox(0,0){\strut{} 400}}%
\put(8886,700){\makebox(0,0){\strut{}}}%
\put(9596,700){\makebox(0,0){\strut{} 600}}%
\put(10196,700){\makebox(0,0){\strut{}}}%
\put(10716,700){\makebox(0,0){\strut{} 800}}%
\put(350,4670){\rotatebox{90}{\makebox(0,0){\strut{}$\lambda_0$}}}%
\put(6462,175){\makebox(0,0){\strut{}$f/[{\rm kHz}]$}}%
\end{picture}%
 

%% file: weaver8.bbl
\begin{thebibliography}{10}
\bibitem{LWcomm}
According to Lobkis and Weaver, that value is more accurate, than the value
given in Ref.~\cite{LobWea03}.
\end{thebibliography}
